# Displacement Current in Classical and Quantum Systems

**David K Ferry[1*], Xavier Oriols[2] and Robert Eisenberg[3]**

[1] *School of Electrical, Computer, and Energy Engineering, Arizona State University, Tempe, AZ 85287 USA*

[2] *Departament d'Enginyeria Electrònica, Universitat Autònoma de Barcelona, 08193 Barcelona, Spain*

[3] *Department of Physiology and Biophysics, Rush University Medical Center, Chicago, IL; Department of Applied Mathematics, Illinois Institute of Technology, Chicago, IL; Department of Biomedical Engineering, University of Illinois Chicago, Chicago, IL USA*

[*]E-mail: ferry@asu.edu

**Abstract**

It is certain that electrical properties—whether slow (sec) or fast (nsec), even optical (fsec)—are described by Maxwell's equations, and there are terms that depend on the rate of change of the electric and magnetic fields. In particular, Maxwell's equation for the curl of the magnetic field contains both the steady current and a term depending upon the temporal derivative of the electric displacement field. The latter is referred to as displacement current, and is generally believed to have been included originally by Maxwell himself, although there is evidence it was earlier considered by Kirchhoff. Maxwell's equations and Kirchoff's circuit laws both are important over the wide range of frequencies with which electronics traditionally deals. And, displacement current is an important contribution to these in both classical and quantum mechanics. Here, the development of displacement current, its importance in both classical and quantum mechanics, and some applications are provided to illustrate the fundamental role that it plays in the dynamics of a wide range of systems.

Keywords: current, gauge invariance, quantum dynamics, classical dynamics

## 1. Introduction

Electrical phenomena always have been interesting to humans. Quite commonly, one refers to an analogy between electrical currents and the flow of water, an analogy described Maxwell himself [1]. This analogy between electrons in a device and water flowing in a tube allows an interpretation of the intricate electromagnetic phenomena in terms of an element familiar to us since our infancy, when playing in the bathtub. But, there is a distinct difference: water is incompressible, while the flow of electricity is quite compressible and leads to the need for Poisson's equation (or its more high frequency versions) [2]. Nevertheless, the analogy is so successful that the movement of electrons inside a device is often referred to as a "flow". From the different flowing regimes, the steady state is the simplest and most common one. However, a steady-state model cannot explain what happens in a waterfall, a whirlpool or when opening a tap, and the same is true in electrical current flow, especially in a semiconductor device. Here, a rigorous and general understanding of the dynamics of electrons and the electromagnetic fields inside devices and circuits, beyond the steady-state regime, requires significant attention to the entire set of Maxwell's equations which incorporate the effects of the displacement current.

Although Maxwell is largely credited with the extension of Ampere's Law, and the introduction of the time-dependent displacement current to his equations in 1861 [3], it is now apparent that Kirchhoff himself published a version that includes displacement current additions to his own d.c. current laws, and this was done *several years earlier than Maxwell* [4,5]. It is important to understand that Maxwell's equations and Kirchoff's circuit laws both are important over the wide range of frequencies with which electronics and electromagnetics traditionally deals. While this historical fact is important to the development of electromagnetics, it is displacement current itself upon which this paper will concentrate.

In the next section, the development of Maxwell's equations, as well as the important role of "gauge" to these equations will be presented. In the following section, the role of Maxwell's equations in classical and quantum dynamics will be explored. Section 4 will address the applications of Maxwell's equations in some areas outside the normal mainstream of electrical activity. Finally, section 5 will outline some conclusions that follow from these discussions.

## 2. Maxwell's Equation and Gauge

The rates of change of charge and electric field are not small in systems that respond on the short-time scale or when the individual fields are large. The mechanisms and properties of current flow vary significantly on the nanosecond (and shorter) time scale. The importance of this clearly lies in Maxwell's extension of Ampère's law, generally considered to be the second of his equations, which may be written as

$$\nabla \times \mathbf{H} = \mathbf{J} + \frac{\partial \mathbf{D}}{\partial t} \qquad (1)$$

and has the corresponding constitutive relations

$$\begin{aligned} \mathbf{D} &= \epsilon_0 \mathbf{E} \\ \mathbf{B} &= \mu_0 \mathbf{H} \end{aligned}, \qquad (2)$$

in the linear case and in the absence of any dielectric polarization, where, $\epsilon_0$ and $\mu_0$ are, respectively, the permittivity and permeability of free space. We return to a discussion of this linearity below.

On the left-hand side of (1), $\mathbf{H}$ is the magnetic field intensity, and for a great many years was measured as so many lines per unit length (in the English system), in keeping with Faraday's lines of force [6]. Today, with the m.k.s system, it is measured as Amps/m. On the right-hand side of (1), the quantity $\mathbf{D}$ is known as the electric flux density, measured as Coul./m$^2$, while $\mathbf{J}$ is the normal particle current density, in A/m$^2$. Thus, one may easily assert that the total current is not only the conduction current $\mathbf{J}$, but must be expressed as

$$\mathbf{J}_{total} = \mathbf{J} + \frac{\partial \mathbf{D}}{\partial t}, \qquad (3)$$

which may be confirmed by taking the divergence of (1) and noting that charge is assured to be conserved with

$$\nabla \cdot \mathbf{J}_{total} = 0 \ . \quad (4)$$

A further constituitive relation, which is merely Gauss' law, is

$$\nabla \cdot \mathbf{D} = \rho \ , \quad (5)$$

where $\rho$ is the charge per unit volume, or charge density. Using this last result in (3) gives us that

$$\nabla \cdot \mathbf{J} + \frac{\partial \rho}{\partial t} = 0 \ , \quad (6)$$

which is the continuity equation for charge and current. The form of the above equations results from Maxwell's own form of the various equations, which he presented in 1861 [1].

It was remarked previously that Kirchoff presented an earlier form of correction to Ampere's Law [4], equation (3) above, which appeared as

$$2\frac{\partial i}{\partial s} = -\frac{\partial e}{\partial t} , \quad (7)$$

where the fact of 2 arises from mid-nineteenth century understanding where it assumed by him that there were two components of moving charge (positive and negative) that would contribute equally to the current. Here, $i$ is the individual particle current, $s$ is position, and $e$ is what Kirchhoff called the "free electricity" (notably the charge to be consistent with the above equations) [4]. It seems to be clear that his form is essentially that of (6).

Let us now turn to Maxwell's first equation, which is commonly expressed as

$$\nabla \times \mathbf{E} = -\frac{\partial \mathbf{B}}{\partial t} \ . \quad (8)$$

which is connected to (1) through the relations (2). Now, the first of two potentials, denoted as the vector potential, is introduced through

$$\mathbf{B} = \nabla \times \mathbf{A} \ , \quad (9)$$

and taking the curl of (8) gives

$$\nabla \times \mathbf{E} = -\frac{\partial}{\partial t}(\nabla \times \mathbf{A}) \ . \quad (10)$$

In order to relate the electric field to the vector potential, an integration constant whose curl is zero may be introduced, and this is used to define the second potential, the scalar potential as

$$\mathbf{E} = -\frac{\partial \mathbf{A}}{\partial t} + \mathbf{C} \ , \quad \mathbf{C} \equiv -\nabla \varphi \ . \quad (11)$$

One normally sees this without the vector potential in our studies of semiconductor devices, but the vector potential term is important, and relates to the connection in high frequency behavior.

How one choses to represent the electric field, and the connections between the vector and scalar potentials, is referred to as a *gauge* condition. To understand how this is invoked, the divergence of (11) is taken to yield

$$\nabla^2 \varphi + \frac{\partial}{\partial t}(\nabla \cdot \mathbf{A}) = -\nabla \cdot \mathbf{E} = -\frac{\rho}{\varepsilon} \ . \quad (12)$$

In fact, if the second term on the left were not present, this would be Poisson's equation commonly used in self-consistent potential solutions in devices. To see how to proceed, (9) and (11) are used in (1) to give

$$\nabla \times (\nabla \times \mathbf{A}) = \\ -\mu_0 \varepsilon_0 \frac{\partial}{\partial t}\left(\frac{\partial \mathbf{A}}{\partial t} + \nabla \varphi\right) + \mu \mathbf{J} \quad (13)$$

These various terms can be rearranged to give a different form as

$$\nabla^2 \mathbf{A} - \mu_0 \varepsilon_0 \frac{\partial^2 \mathbf{A}}{\partial t^2} = -\mu \mathbf{J} \\ + \nabla\left(\nabla \cdot \mathbf{A} - \mu_0 \varepsilon_0 \frac{\partial \varphi}{\partial t}\right) \quad . \quad (14)$$

Here, the terms on the upper line compose a wave equation for the vector potential in which the latter is driven by the particle current density. It is the lower line that adds terms that change the equation. This is where gauge appears to play a role. In order to have this wave equation, and to also have a wave equation for the scalar potential, it is convenient to set the term in parentheses to zero. This condition is known as the *Lorentz gauge*, or sometimes simply as the *gauge equation*

$$\nabla \cdot \mathbf{A} - \mu_0 \varepsilon_0 \frac{\partial \varphi}{\partial t} = 0 \ . \quad (15)$$

This may now be used in (12) to give an equivalent wave equation for the scalar potential

$$\nabla^2 \varphi - \mu_0 \varepsilon_0 \frac{\partial^2 \varphi}{\partial t^2} = -\frac{\rho}{\varepsilon}, \quad (16)$$

in which the solutions of the scalar potential are driven by the charge density. Here, it is clear that Poisson's equation is only a low frequency approximation to this wave equation!

*2.1 Gauge and Gauge Invariance*

The Lorentz gauge described above is just the beginning of possibilities. There are further constraints that can be imposed. For example, a further approximation is to invoke the Coulomb gauge, or the electrostatic gauge as it is sometimes called, in which

$$\nabla \cdot \mathbf{A} = 0 \quad (17)$$

is invoked. This constraint, together with the Lorentz condition (15), requires

$$\frac{\partial \varphi}{\partial t} = 0 \ . \quad (18)$$

Now, the wave equation (16) reduces to the normal Poisson equation used in situations when a self-consistent potential must be found. This familiar result arises from a choice of gauge; it is neither automatically true nor basic. Hence, when one solves the Poisson equation for a device, an assumption is being made that only low frequency effects are of interest; the potential and electric field instantaneously follow variations in charge. This assumption is also followed by the assumption that the electric field is found solely from the scalar potential, thus ignoring the time derivative of the vector potential in (11). This further results in the assumption that the charge density is *static and time invariant* on the time scale of interest. This means that the scalar potential follows the charge density change instantaneously, in clear violation of relativity. So, if one wants to use the Coulomb gauge in device simulation, it must first be determined that any charge changes, due to the imposition of self-consistency, must be slow enough to validate use of this gauge, and care must be exercised in evaluating fields and potentials.

Another common gauge arises in the study of the magnetic field effect especially when quantum mechanical effects are being studied. There are two gauge choices that are made, both of which are consistent with the Coulomb gauge discussed in the previous paragraph. These arise from the manner in which the magnetic field and the vector potential are constructed to satisfy equation (9). One of these is termed the *Landau gauge* [7], which is described through the form

$$\mathbf{A} = \pm B x \mathbf{a}_y \ . \quad (19)$$

This leads to a magnetic field in the positive or negative *z*-direction. The second approach uses what is termed the symmetric gauge

$$\mathbf{A} = \frac{1}{2}\left(-B y \mathbf{a}_x + B x \mathbf{a}_y\right) \ . \quad (20)$$

This form is particularly useful as it carries the quantization of the states in the magnetic field into the relavistic regime [8].

The use of a gauge remains important in that it connects the magnetic fields and the electric fields. This connection requires that the fields satisfy the Lorentz condition (15). When this is true, the fields are said to be *gauge invariant*. If the fields do not satisfy this requirement, then the full equations (12) and (14) must be solved to give the two potentials self-consistently. Given an initial gauge, a gauge transformation can be made that is subject to the new potentials as

$$\begin{aligned} \mathbf{A} &\to \mathbf{A} + \nabla \Lambda, \\ \varphi &\to \varphi - \frac{\partial \Lambda}{\partial t} \ . \end{aligned} \quad (21)$$

The function $\Lambda$ provides the transformation that creates the change in gauge.

It is important that, in general, (21) gives the same electromagnetic fields, which implies that the same forces and dynamics are generated by fields in the new gauge as were generated by the original fields. Moreover, (21) is a major gauge change for time-varying fields. There are simple gauge changes which are not time varying, and which essentially are decisions on how to describe energy and potential, as well the fields independently. It is important to recognize that Newton's law, which for charged particles depends on the electric field, does not change with the gauge. Therefore, one cannot know which form of electromagnetic potentials (before or after the gauge transformation) have generated the dynamics of a given experiment. The fact that the gauge is unobservable does not mean that the fields are unobservable. Rather, the gauge condition implies that the observed fields are gauge invariant.

In both classical mechanics and quantum mechanics, the electromagnetic interactions may be taken into account through a change in the momentum, known as the Peierls' substitution, through

$$\mathbf{p} \to \mathbf{p} - e\mathbf{A} . \qquad (22)$$

This is particularly useful in Hamiltonian dynamics, and particularly quantum mechanics, in that it requires the vector potential to satisfy any uncertainty relations. In order to keep the wave function gauge invariant (Newton's law or the Schrödinger equation have to keep their same shape in any gauge), apart from the transformation (21) in the electromagnetic potentials, the wave function must transform as follows [9]

$$\psi(\mathbf{r}) \to e^{ie\Lambda/\hbar}\psi(\mathbf{r}), \qquad (23)$$

with $\Lambda$ being the local (space-time dependent) function used in (21). Of course, the observable results obtained from the wave function are independent of the gauge transformation (for example the probability $|\psi(\mathbf{r})|^2$ is independent of $\Lambda$ by construction).

The computation of the wave function in one specific scenario can be more easily formulated in one gauge than in another. For example, there is a common gauge transformation when the wavelength of the electromagnetic field is much larger than the region of interest where the system's dynamics are described. Thus, the spatial dependence of the vector potential can be neglected in that region, $\mathbf{A}(\mathbf{r},t) \approx \mathbf{A}(\mathbf{r_0},t)$. Then, the following gauge $\Lambda(\mathbf{r},t) = -\mathbf{r} \cdot \mathbf{A}(\mathbf{r_0},t)$ when applied to (21) gives a zero vector potential $A \to A + \nabla \Lambda = A(\mathbf{r_0},t) - A(\mathbf{r_0},t) = 0$ and a new scalar potential $\varphi \to \varphi - \frac{\partial \Lambda}{\partial t} = \varphi(\mathbf{r},t) + \partial(\mathbf{r} \cdot \mathbf{A}(\mathbf{r_0},t))/\partial t$. In other words, when this gauge transformation is applicable, the shape of the Schrodinger equation remains the same but the vector potential term $e\mathbf{A}$ in (22) disappears at the price of dealing with a more complicated scalar potential.

An important aspect of this last equation arises when the electromagnetic fields themselves are quantized. Generally, this latter process involves expanding the fields in their respective Fourier modes, which may be represented as a sum over harmonic oscillators (one oscillator per mode). This quantization of the fields is generally described as second quantization, while normal quantization of a harmonic oscillator is first quantization. The difference is mostly semantic. The important aspect arises from the fact that the fields themselves must be gauge invariant. Therefore, any quantization of the fields also must require the corresponding observable results to be gauge invariant. An example of such gauge invariance in quantum transport may be found in the use of nonequilibrium Green's functions [9,10].

A final remark is that the gauge transformation actual requires the existence of the displacement current and a continuity equation. This is discussed further in the Appendix.

*2.2 Polarization and Magnetization*

Polarization and magnetization describe the processes by which various materials modify the propagation properties that they exhibit. That is, they describe effects which change the constitutive relations (2). These changes can be linear effects or extremely nonlinear effects. The generalized form for these two equations becomes

$$\begin{aligned}\mathbf{D} &= \epsilon_0\mathbf{E} + \mathbf{P}\\ \mathbf{B} &= \mu_0\mathbf{H} + \mathbf{M}\end{aligned}. \quad (24)$$

Here, **P** is the polarization and **M** is the magnetization. That is, the polarization represents additional charge in the material and the magnetization represents additional sources of magnetic flux.

As commented above, the properties of materials can lead to linear effects or quite nonlinear effects. In the linear case, the polarization is represented as a simple function of the electric field **E**, while the magnetization is represented as a simple function of the magnetic field intensity **H**. In this linear response approach, (24) is then written as

$$\begin{aligned}\mathbf{D} &= \epsilon_0\mathbf{E} + \chi_e\epsilon_0\mathbf{E} = \epsilon\mathbf{E},\\ \mathbf{B} &= \mu_0\mathbf{H} + \chi_m\mu_0\mathbf{H} = \mu\mathbf{H},\end{aligned} \quad (25)$$

where the $\chi's$ are the susceptibilities. In this form, the "relative" dielectric constant $\epsilon_r = 1 + \chi_e$, and "relative" permeability $\mu_r = 1 + \chi_m$, have been introduced. Then, the total permittivity and permeability are the products of these relative values and the free space values. This approach is simplistic, and it neglects a great many effects that are present in real materials. This is because the polarization and magnetization are really quite complicated entities and vary differently from simple single material crystalline atomic materials to quite difficult organic and biological materials. Even such a common material as water does not have fully understood polarization behavior.

In a linear material such as a normal semiconductor like silicon, the electric susceptibility is determined by the polarization of the bonding (outer shell) electrons that form the covalent bond. At frequencies up to the extreme ultraviolet, this susceptibility is constant at a value of 10.68 in the linear response regime. In compound semiconductors, however, there can be an atomic contribution to the susceptibility due to the slight ionic contribution to the bonding. This atomic contribution appears for microwaves below around 20-40 micrometer wavelength. Hence, there is an optical value and a larger "low frequency" value. In gallium arsenide, for example, the two values for the susceptibility are 9.89 and 10.9, with the transition occurring in the region 32-24.4 micrometer. A more complicated material such as $SiO_2$ has two such transitions, with three different values for the susceptibility [11]. Liquids in general, and ionic solutions in particular have much more complicated behavior [12]. Life tends to exist in ionic solutions, for the most part. Hence, biological material can be expected to be considerably more complicated in their dielectric response.

However, such a simple material as silicon can be quite nonlinear in reality [13]. In such a case, the polarization can be expanded as

$$\begin{aligned}P(t) = \varepsilon_0[&\chi_{e0}E(t)\\ &+\chi_{e1}E^2(t) + \chi_{e2}E^3(t)+\dots]\end{aligned}. \quad (26)$$

This is especially useful in nonlinear optics, where silicon can be used as a waveguide material and takes advantage of the extensive technology for such processing. The first-order process $(\chi_{e0})$ is the simple dipole contribution discussed in the previous paragraph. The real part of this is the normal propagation, while the imaginary part describes gain or loss proceeses in the material (to be described further below). The second-order process ($\chi_{e1}$) is not present in materials such as silicon which are inversion symmetric. The third-order process ($\chi_{e2}$) can give rise to third-harmonic generation or four-wave mixing in which two incoming frequencies (or two photons at the same frequency) parametrically combine to create photons at two different frequencies [13]. The latter is an important process in quantum optics, especially in quantum information processing. The simple idea of a constant permitivitty is no longer allowable even in this simple material.

Then there are polarizations and magnetizations that are relatively independent of **E** and **H**. These are materials like ferroelectrics and ferromagnetics, and usually involve iron (hence the names) in some form. While in some cases, these polarizations and magnetization can be reversed in sufficiently high fields, which gives rise to hysteresis in **D** vs. **E** and **B** vs. **H** curves, there are known materials which this is not the case, and the two effects are basically permanently fixed in direction. One such type of material is a pyroelectric which possesses a permanent polarization.

*2.3 Conduction and Displacement Currents*

Equation (3) defines the total current in a simple and complete manner, since it describes the continuity of charge at a point. That is, as may be seen in (6), the net conduction currents flowing into a point in space must provide a temporal change in the amount of charge existing at that point. The simplest example of this is the common capacitor, such as shown in figure 1. In this case, the insulator that separates the two parallel plates is vacuum, but it could be any good insulator. The common circuit law is

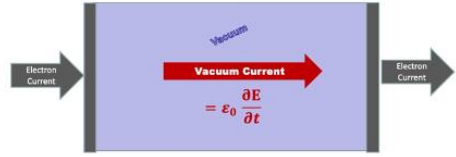

**Figure 1**. A vacuum capacitor. In a normal capacitor, the permittivity would be corrected with the relative permittivity of the material used as dielectric in the vacuum region.

that the a.c. current through the capacitor is given by

$$I = C \frac{dV}{dt}, \qquad (27)$$

where *V* is the applied voltage and

$$C = \frac{\epsilon A}{L}, \qquad (28)$$

where *A* is the area of the capacitor plates and *L* is the distance between the two plates. By converting the current to current density with $J = I/A$, and the voltage to electric field with $E = V/L$, the current density is shown to be entirely a displacement current

$$J = \epsilon \frac{dE}{dt}. \qquad (29)$$

Thus, the conduction current that flows into, and out of, the capacitor through the wires, actually flows *through* the capacitor by displacement current. Current continuity is maintained throughout the system.

This can be extended into the general a.c. current regime by assuming that the currents and fields vary as $exp(i\omega t)$. With this, and using Maxwell's constitutive equations, one observes that

$$\mathbf{J}_{total} = \sigma \mathbf{E} + i\omega\epsilon \mathbf{E}. \qquad (30)$$

Hence, it is obvious that the displacement current is a key part of the impedance relationship between field and current. This impedance follows directly from (3) as

$$Z = \sigma + i\omega\epsilon. \qquad (31)$$

(Note that both $\sigma$ and $\epsilon$ are the total functions; e.g., $\epsilon = \epsilon_r \epsilon_0$, as discussed above.) This is a clear link between circuit theory (no matter the frequency) and electromagnetic theory, which is crucial to the entire field of electronics. The importance of this impedance directly affects modern semiconductor devices, as the carriers (electrons and hole) in these devices have a natural inductive behavior in addition to their normal conductance. This point will appear again below.

Now, one sees that Kirchoff's circuit equations and Maxell's electromagnetic equations are intimately connected, and express the same physics in different environments. No matter the electronics system, these equations, and displacement current, are essential to the full understanding of impedance in electronics.

*2.4 A Simple Gauge Example. I*

It might occur to the reader that pure ballistic transport should yield infinite conductance. That is, if there is no scattering, what causes resistance that gives rise to only a finite conductance. Here, one has to consider the transition between the metallic leads and the channel which itself can lead to non-zero resistance. This is a contact resistance between the ballistic channel and the reservoirs which give access to it. This can be examined with a simple example that also illustrates the possibility of a gauge transformation.

Consider first the case in which the potential energy is uniformly zero throughout the system, and an electric field is applied in the region between $x = 0$ and $x = L$. This electric field may be written as a vector with the form $\boldsymbol{E} = -E\boldsymbol{a}_x$, where $E$ is the amplitude of the field. Particles which enter the active region at $x = 0$ are then accelerated in the field according to Newton's law as

$$m\frac{dv}{dt} = eE$$
$$v = \frac{eE}{m}t + v_0 \qquad (32)$$

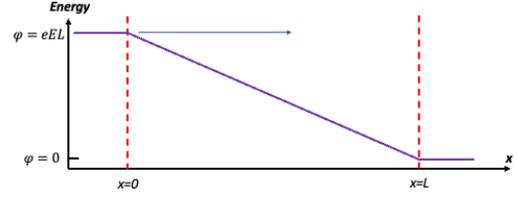

**Figure 2.** Variation of the energy for a situation in which the electric field is gauged in with the potential.

For ease of following the train of thought, the integration constant $v_0 = 0$. A second integration gives the position of the particle as

$$x = \frac{eE}{2m}t^2 . \qquad (33)$$

The time to traverse the length L is easily found to be

$$t = \sqrt{\frac{2mL}{eE}} . \qquad (34)$$

From, this one can find the velocity of the particle at the end of the active region to be

$$v = \sqrt{\frac{2eEL}{m}} , \qquad (35)$$

and this leads to the (kinetic) energy gain as

$$T = \frac{1}{2}mv^2 = eEL . \qquad (36)$$

As an alternative approach, a gauge transformation is made in which the electric field is transferred to the total energy, so that the latter consists of both potential and kinetic energy. With this alternative, the energy appears as shown in figure 2. Now, the particles entering the active region at $x = 0$ travel ballistically (at constant energy as shown with the blue arrow) across the active region. That is, they move in a manner that conserves the *total* energy, the sum of the kinetic and potential energies. As they move, potential energy that exists at the entrance (on the left) is gradually converted to kinetic energy, so that the energy gain at $L$ is still $eEL$. That is, the physics of the energy and velocity gain is precisely the same in both gauges. Hence, the gauge transformation is merely a method of looking at the problem in different ways in order to find one approach that explains the problem in a perhaps better manner.

The problem with this simple example is that Kirchhoff's laws are violated. The conduction current density in either situation is

$$J = -nev = \rho v, \quad (37)$$

where $n$ is the number of particles per square meter in the cross-section of the device, and $\rho$ is the charge density. Since the velocity increases as one moves from left to right, the particle density $n$ must decrease in order to have current continuity. This simple fact has not been considered in these discussions so far, but it must be considered in order to satisfy Kirchhoff's current law (4). Thus, either approach above is inconsistent with (4). In order to make it consistent, Poisson's equation must be used.

To proceed, the first gauge will be used, so that the initial boundary conditions can be set to

$$\begin{aligned}\varphi(0) &= 0 \\ \left.\frac{d\varphi}{dx}\right|_{x=0} &= 0\end{aligned} \quad (38)$$

Poisson's equation is then

$$\frac{d^2\varphi}{dx^2} = -\frac{\rho(x)}{\epsilon} = \frac{J}{\epsilon v(x)}, \quad (39)$$

and (37) has been used. Using (35), the velocity can be related to the potential as

$$v(x) = \sqrt{\frac{2e\varphi(x)}{m}}, \quad (40)$$

and (39) can be rewritten as

$$\frac{d^2\varphi}{dx^2} = -\frac{1}{\epsilon}\sqrt{\frac{m}{2e}}\frac{J}{\sqrt{\varphi}}. \quad (41)$$

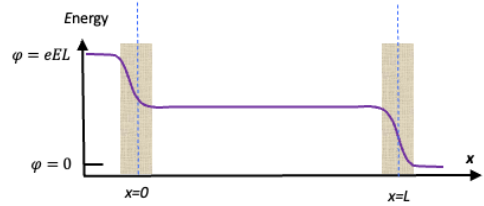

**Figure 3.** Potential variation for ballistic transport and a result satisfying Kirchhoff's current law. The shaded areas, in which the potential drops occur, are contact resistances. To support the potential drops, these regions must have dipole charges composed of accumulation and depletion of charge.

The integrations are relatively straight forward [14] and yield the result that

$$J = \frac{4\epsilon}{9L^2}\sqrt{\frac{2e}{m}}\varphi^{3/2}. \quad (42)$$

This relatively well-known result is the Langmuir-Child Law [15,16]. It is clear here that the conductance is not linear in potential. While this result holds for ballistic transport, it is not the only possible result, in particular in semiconductors with a relatively constant doping density.

In a semiconductor, the density is usually set by the doping. When the number of carriers drops, such as is required in the first two cases due to the need to satisfy Kirchhoff's current equation (4), an additional space charge is created in the channel due to the un-neutralized donors. The result of this argument is that the linear potential drop shown in figure 2 cannot exist, as pointed out above. if the carriers are moving via ballistic transport. Such a linear potential drop only can occur if there is sufficient scattering to assure that the carriers move with a near-equilibrium energy. The logical conclusion that allows for constant density is that the electric field must essentially be *near zero* in the actual channel, if the carriers are to move via ballistic transport.

This implies that the potential drop must divide between the cathode and the anode transition regions, the "magic" regions discussed above that were referred to as contact resistance regions. Such a potential variation is illustrated in figure 3. It is usually assumed that most of the discontinuity is at the cathode, but not always. Now, ballistic transport can occur through the constriction without the carriers gaining excess energy from the applied bias. At the same time, the potential drops in the transition regions now require that dipole charge densities exist at each transition. The potential drop can only occur through the existence of such dipole charge densities. This complication of the simple example arises from the additional need to conserve current in the "device" that is being considered. Both solutions, figures 2 and 3, are mathematically possible, but once current conservation is considered, only figure 3 remains. The physical rationale is that the desire for pure ballistic transport cannot over-rule over physical constraints, such as current continuity which is present in Maxwell's equations through (6). The physically pleasing example of figure 2 just doesn't satisfy (6), so must be excluded from the discussion.

*2.5 A Simple Gauge Example. II*

As a second example of a simple gauge transformation, consider the standard harmonic oscillator that is treated in most quantum mechanics textbooks. For this system, the Schrödinger equation may be written as

$$-\frac{\hbar^2}{2m}\frac{d^2\psi}{dx^2} + \frac{m\omega^2 x^2}{2}\psi = E\psi . \quad (43)$$

Here, ω is the natural frequency of the oscillator, and the other symbols have their normal meaning. Referring to a textbook [17], the energy of the oscillator is quantized into Planck units as

$$\mathcal{E}_n = \left(n + \frac{1}{2}\right)\hbar\omega , \quad (44)$$

where *n* is an integer. The wave functions form an infinite set, in which each member may be defined by

$$\psi_n(x) \sim exp\left(\frac{-m\omega x^2}{2\hbar}\right) H_n\left(\sqrt{\frac{m\omega}{\hbar}}x\right), (45)$$

within a normalization constant, and $H_n$ is a Hermite polynomial.

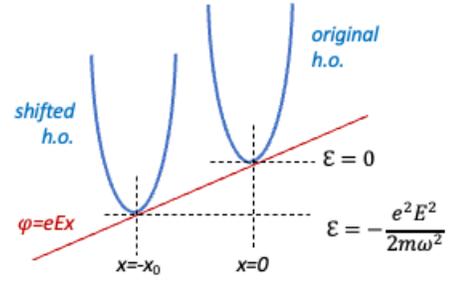

**Figure 4.** In an applied electric field, the harmonic oscillator is shifted both in position and energy along the electric potential line, shown in red.

Now, consider the addition of an electric field through the. potential $\varphi = eEx$. The standard approach is to treat this potential as a small perturbation, and to use perturbation theory to examine how the energy levels and wave functions change. At first order, the perturbation leads to the coupling of each state to its neighbors above and below in energy, while the second-order produces a downward correction for each level of

$$\Delta\mathcal{E} = -\frac{e^2 E^2}{2m\omega^2} . \quad (46)$$

But, these results are obtained only at the expense of considerable calculations.

To understand both (44) and the above statement that the field couples a level with the ones just below and just above, one can note that the Hamiltonian on the left side of (43) can be split into the product of two operators as [18]

$$a = \sqrt{\frac{m\omega}{2\hbar}}\left(x - \frac{\hbar}{m\omega}\frac{\partial}{\partial x}\right),$$
$$a^\dagger = \sqrt{\frac{m\omega}{2\hbar}}\left(x + \frac{\hbar}{m\omega}\frac{\partial}{\partial x}\right). \quad (47)$$

A little manipulation shows then that the Hamiltonian can be written as

$$H = \left(a^\dagger a + \frac{1}{2}\right)\hbar\omega = \mathcal{E}. \quad (48)$$

These two operators raise and lower the energy state as

$$\begin{aligned} a^\dagger \psi_{n-1} &\to \psi_n \\ a\psi_{n+1} &\to \psi_n \end{aligned}, \quad (49)$$

where a number of normalization parameters have been left out. But, using the product of operators in (48) just gives the original state and the number of particles and this leads to (44). When one now turns to the perturbation change in the wave number, one notes that the position operator $x$ in the potential $\varphi$ leads to

$$x \sim a + a^\dagger, \quad (50)$$

leads to the perturbation relation

$$\delta\psi_n \sim \langle\psi_n|(a + a^\dagger)|\psi_m\rangle \quad (51)$$

in Dirac notation, and (49) tells us that $m = n \pm 1$. Thus, the change is a particular state comes only from coupling to the ones above and below.

It turns out that the solution with the electric field is much easier to obtain with a gauge transformation. It is not obvious that this is the approach being used, but it will become evident at the end. Here, the additional term in the Hamiltonian is taken together with the harmonic potential and squared as

$$\begin{aligned} V &= \frac{m\omega^2 x^2}{2} + eEx \\ &= \frac{m\omega^2(x+x_0)^2}{2} - \frac{e^2 E^2}{2m\omega^2} \end{aligned}. \quad (52)$$

It is clear that the energy shift applies to each and every eigenstate, since it does not involve any individual energy. Moreover, the first term in the second line gives the information that the harmonic oscillator is shifted in position by the amount $x_0$, where

$$x_0 = \frac{eE}{m\omega^2}. \quad (53)$$

Hence, the gauge transformation arises from using the field in the potential itself, which results in a physical shift of the harmonic oscillator as shown in figure 4. The new wave function is found by using (45) with $x$ replaced by $x + x_0$. This can be seen by introducing the displacement operator to shift the wave function as

$$\psi(x + x_0) = e^{ipx_0/\hbar}\psi(x). \quad (54)$$

Now, the gauge potential can be found by comparing this result with (23), so that the displacement operator actually is a gauge transformation. Now, this gauge shifts the wave function in position rather than in momentum. The pre-factor in (54) is known in quantum mechanics as a displacement or translation operator [17]. If one expands the exponential term, the full Taylor series for the shifted wave function (in terms of the unshifted one) is obtained.

## 3. Some Implications

The two simple examples for gauge transformations in the previous sections illustrated the effect for a classical particle and a quantum wave function. But, is there a difference? Certainly, classical mechanics has a long history of using particles which are subjected to Hamilton's equations of motion or Lagrangian mechanics [19]. But, many readers may not be familiar with the use of particles in quantum mechanics, although connection with atomic physics should clearly show the possibility of particle contributions. Indeed, Feynman clearly had particles in mind in constructing his path integral approach to quantum mechanics [20]. Already in 1928, Kennard [21] had shown that quantum particles would move in response to the classical forces plus a quantum force of the form of the Bohm potential [22]. And, in most cases, quantum motion and wave functions, either directly or through simulations of Wigner functions or density matrices, are amenable to particle treatments [23]. The reader should keep in mind that particles and waves are parallel approaches to a quantum mechanics that contains both [24].

*3.1 Guided Waves*

The importance of the displacement current is well recognized today in both electromagnetics and circuits, particularly for the a.c. case. In many cases, however, the treatment reduces to linear response in which a single relative permeability or relative permittivity is used. Linear response is a very special case, and represents only a very small part of the world of electromagnetics. Certainly, the general case is that **P** is time varying, just as **E** is. If **P** is also either nonlinear in the field or is inhomogeneous, the result is not simple wave propagation, but can lead to very complicated nonlinear equations and/or distinctly different propagation properties in different crystalline directions (within a crystalline material) [25]. The entire field of nonlinear optics depends upon moving beyond linear response. The fact that there is so much effort (and publications) in microwave theory and techniques unfortunately masks the point that it is based upon a relatively simplistic approximation.

A particular example of the difficulties is the millimeter integrated circuit (MMIC). Transport of the millimeter waves on the MMIC is usually by strip lines (open waveguides which induce propagating waves guided by a top surface metal strip line and the underlying ground plane, using a non-absorbing substrate material), although coplanar waveguides are also used [26].

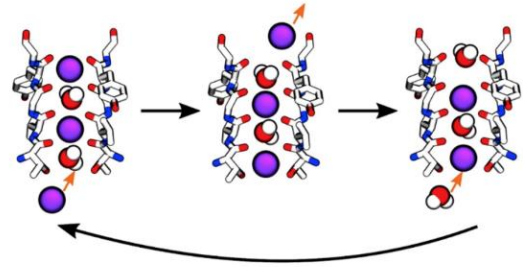

**Figure 5**. Serial passage of potassium ions through an ion channel. The black just denotes repeating the sequence of conformations. Reprinted from A. Mirenenko *et al.* [31], under the creative commons usage license.

These waveguides have relatively low impedance (lower than free space), but must be matched to the very-high impedance (reactive in most cases) inputs and outputs of the transistors. That is, the inputs are largely dominated by the gate capacitance. This requires complicated matching networks to be included in the circuit. Moreover, the millimeter waves must be isolated from the power leads, and the d.c. power must be isolated from the waveguides. All of this requires design constraints which are not always compatible to each other. Generally, this leads to a difficult design problem [27].

*3.2 Ion Channels*

An important application of the conservation of total current occurs in the ion channels of biological membranes. These ion channels are mostly narrow pores through proteins that allow otherwise impermeable ions to pass into cells. Ion channels control an enormous range of biological functions in health and disease and are extensively studied. The narrow pores of biological channels are rarely wide enough to allow ions to pass by each other with high probability. The current flow through the pores has been viewed as a single file hopping phenomenon [28,29]. The ion channels of nerve, skeletal, and cardiac muscle responsible for nerve signaling and the coordination of contraction use total current to make the nerve signal, as is apparent from both experiments and theory [30]. It is important to recognize that nerve signals are not d.c., but are pulsed a.c., and require displacement current for their efficient signal propagation.

The single file passage of the ions certainly is of great importance for the charge current carried by these ions through the channel, as shown in Fig. 5 [31]. But the total current through the pore of the channel includes another component, the displacement current produced by the polarization of matter and space. The sum of those components is conserved even though the individual components are not. The charge current varies dramatically with position. The displacement current varies dramatically with position. But, their sum does not vary with position, at all, as indicated by (6).

Maxwell's equations—and versions of Kirchhoff's law that are consistent with these—ensure that total current is conserved whenever these equations are used [32]. In a narrow single file channel, the displacement current takes over from the charge current (and vice versa) exactly so that the total current is constant along the length of the narrow channel. The consequences of the interplay of charge current and displacement current is to simplify the system dramatically. The total current does not vary with spatial location in a narrow one dimensional system. However complicated the hopping and single file behaviors actually are, the total current is the same at all spatial locations in the channel because one component of the total current takes over from the other, to make it so. The electric and magnetic fields change the movement of charges on the atomic scale according to the requirement of electromagnetics.

The implications for atomic scale theory were clearly known in theories of one dimensional transport [32]. In other words, a theory of the total current does not need to have the spatial location as an independent variable. Of course, a theory of total current is not a complete theory of electrodynamics, let alone charge movement. The spatial variable is obviously needed for complete understanding. In many situations, however, a measurement of total current is enough to allow significant understanding and control of a system. Those situations include many of the circuits of our electronic technology. They also include many ion channels.

### 3.3 High Frequency Quantum Devices

The great success of our information society is based on encoding the physical values of currents and voltages inside electron devices as digital (or analog) information. Typically, the simulation of such devices is done considering only the particle current, while ignoring the displacement current. But, this fails at high frequency as noted above. To understand when this low frequency assumption is acceptable, consider some values for the total current $\mathbf{J}_{total}$ mentioned in (3). On the right-hand side of (3), the particle current is proportional to the electric field through the conductivity $\sigma$ and the displacement current is evaluated assuming a sinusoidal temporal dependence of the displacement field $\mathbf{D} = \epsilon \mathbf{E}$ corresponding to a frequency ω. The typical conductivity $\sigma$ in Silicon is less than $10^{-1}\,\Omega^{-1}m^{-1}$ and, using $\epsilon = 10\epsilon_0 = 8.85 \times 10^{-11}\,\mathrm{F}m^{-1}$, one finds the displacement current in semiconductors devices can be safely ignored up to few GHz. However, since the displacement current in (3) grows linearly with ω, the displacement current cannot be ignored for high enough frequencies (obviously, the displacement current cannot be ignored at any frequency in capacitors because the conductivity is zero, as discussed above).

A relevant question is just how is displacement current modeled in semiconductor devices operating at tens of GHz? A straight-forward answer comes from the semi-classical simulation of electron devices. For example, the typical Monte Carlo solution of the Boltzmann equation provides the semi-classical trajectory $x(t)$ for each electron so that the total charge density $\rho$ can easily be defined. Then, the displacement current in (29) can be evaluated from the time-derivative of **D** obtained by using Gauss' Law in (5). But, when quantum phenomena become relevant, the mandatory inclusion of the displacement current in quantum transport simulators becomes a more complicated issue, either from a computational or fundamental point of view [33,34].

According to the orthodox quantum theory, any measured property of a system coincides with the eigenvalue of an operator linked to such property, and the state of the system "collapses" into the eigenstate of such eigenvalue. For modeling d.c. currents, the "collapse" is ignored assuming that time-averaged current is equivalent to an average over identical devices whose current is measured just once. However, the previous ergodic argument is no longer valid in far-from-equilibrium semiconductor devices, especially at high frequencies. In principle, one would have to face the perplexing effects of the "collapse" postulated by the orthodox machinery. However, in practice, this orthodox theory is avoided by more causal versions of quantum mechanics [33]. And, in these approaches, the high frequency performance of quantum devices is mainly understood from static (d.c.) quantum simulations. It is assumed that the quantum device behaves as a (small-signal) circuit. The resistances and capacitances of such circuit are then computed from static (d.c.) quantum simulations to evaluate variations of current (conductance) or charge (capacitance) for different voltage.

Fortunately, a direct quantum modelling of the displacement current in quantum devices without either the (small-signal) circuit assumption or the perplexing effects of the "collapse" law is possible. There are quantum theories where electrons have well-defined properties independently of their measurement (observation). Such quantum theories without observers, for example Bohmian mechanics [35,36], are well known in the community dealing with the foundations of quantum mechanics. These remain mostly ignored in the electron device community. Yet, the great advantage of the Bohmian formulation of quantum phenomena is that the evaluation of the displacement current can be done following a similar strategy used in the semi-classical Monte Carlo simulations [13,37,38]. More details about such Bohmian trajectories will be presented in Sec. 4. Once quantum (Bohmian) trajectories $x(t)$ are determined, satisfying the continuity equation (6), the computation of the displacement follows in a straight forward manner without any need of the orthodox "collapse". In Figure 6, the total current (particle plus displacement current), computed from Bohmian trajectories, is plotted as a function of time for a resonant tunneling diode biased on a sinusoidal signal at different frequencies. The frequency-dependent non-linear behavior of the total current can be related to memory effects. There is plenty of room for unexplored applications of tunneling devices working at frequencies higher than the inverse of the electron transit time, where displacement current becomes more important than the particle current [39].

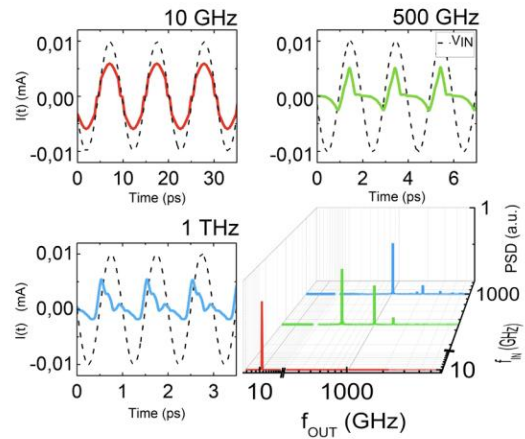

**Figure 6.** Total current for a resonant tunneling diode as a function of time for different input frequencies $f$ of a small-signal input voltage (dashed black in arbitrary units). In bottom left, power spectral density (PSD) as a function of the output frequency for the three currents, confirming (nonlinear) harmonic generation [39].

In order to be fair, it is necessary to point out that many in the quantum community. prefer to use the Schrödinger equation based non-equilibrium Green's functions [40] to simulate the full nonlinear response of semiconductor devices [41]. Normally, this is difficult and consumes considerable computational resources, and does not yield the efficient computations using Monte Carlo approaches [42]. More recently, progress has been made in using Monte Carlo techniques to evaluate the Green's functions and greatly speed up device simulations [43].

## 4. Displacement Current in Quantum Mechanics

As explained in the above sections, the interaction of a classical particle with a classical electromagnetic field is modelled using the so-called canonical momentum $p = mv + eA$ in the definition of the kinetic energy $mv^2/2$. Then, the typical Hamiltonian becomes

$$H = \frac{1}{2}mv^2 + e\varphi = \frac{(p-eA)^2}{2m} + e\varphi. \quad (55)$$

The description of a classical state in a given experiment is a trajectory $r^j(t)$, where the superscript $j$ indicates one of a large number of possible trajectories. To understand displacement current in a quantum scenario, one must explain how the classical Hamiltonian (55) is transformed into a quantum one. To simplify the discussion, only one electron will be considered in this section. Initially, it may not be obvious to the reader why this solitary electron does would not simply move freely, without interaction. The electromagnetic fields in (55), represented by the vector $A$ and scalar $\varphi$ potentials (see Sec. 2), are generated by other particles. However, to avoid dealing with a many-particle scenario that will unnecessarily complicate all discussion, only one electron interacting with the electromagnetic field is considered. In particular, a system with a quantum electron interacting with a classical electromagnetic field (such as light) will be presented first. Then, a system with a quantum electron and quantum light will be discussed. A numerical example will be provided in this latter case.

### 4.1 Quantum Electron and Classical Light

The quantum version of the Hamiltonian (55) is obtained through the so-called canonical quantization [23,44], where the classical canonical variables $r$ and $p$ are substituted by quantum non-commuting operators, $r \to \hat{r} = r$ and $p \to \hat{p} = -i\hbar\nabla$ with $[\hat{r}, \hat{p}] = i\hbar$ (this expression is understood as being zero unless equal components of the two vector operators are considered; that is $[\hat{r}_i, \hat{p}_j] = i\hbar\delta_{ij}$). Then, the Hamiltonian (55) can be written as

$$H_e = \frac{(-i\hbar\nabla - eA)^2}{2m} + e\varphi. \quad (56)$$

The quantum state is defined by the (complex) wave function $\Psi(r,t)$, which is a solution of the following Schrödinger equation

$$i\hbar\frac{\partial \Psi(r,t)}{\partial t} = H_e\Psi(r,t).$$
$$= \left(\frac{(-i\hbar\nabla - eA)^2}{2m} + e\varphi\right)\Psi(r,t) \quad (57)$$

Since quantum mechanics is a statistical theory, $\Psi(r,t)$ is not a description of a single experiment, but a description of the ensemble of all (identical) experiments. The differences between displacement current linked to an ensemble of experiments and the displacement current for a single classical experiment will be emphasized here.

Once the wave function $\Psi(\mathbf{r},t)$ is obtained from (57), the "charge" density may be determined from it as

$$\rho(\mathbf{r},t) = e|\Psi(\mathbf{r},t)|^2 . \quad (58)$$

The physical meaning of $\rho(\mathbf{r},t)$ is the probability of finding the electron at position $\mathbf{r}$ at time $t$ when the same experiment is repeated many times [23,45]. It is important to note that a continuity equation is "hidden" in (57). The temporal variations of the charge in (58) are computed as

$$\begin{aligned}\frac{\partial \rho}{\partial t} &= e\Psi \frac{\partial \Psi^*}{\partial t} + e\Psi^* \frac{\partial \Psi}{\partial t} \\ &= e\Psi \frac{H_e \Psi^*}{-i\hbar} + e\Psi^* \frac{H_e \Psi}{i\hbar}\end{aligned}, \quad (59)$$

and this expression leads to a continuity equation

$$\frac{\partial \rho(\mathbf{r},t)}{\partial t} + \nabla \cdot J(\mathbf{r},t) = 0 , \quad (60)$$

where the conduction (or particle) current $J(\mathbf{r},t)$ is defined as

$$J(\mathbf{r},t) = \frac{e}{m} Im\{\Psi^*(\mathbf{r},t) \\ \times \left(\nabla - \frac{e}{\hbar}\mathbf{A}(\mathbf{r},t)\right)\Psi(\mathbf{r},t)\} . \quad (61)$$

The terms $\rho(\mathbf{r},t)$ and $J(\mathbf{r},t)$ are the quantum sources of the electromagnetic fields through (1) and (8) and these must satisfy Maxwell's equations. In particular, Gauss' law relates the charge density to the electric field as follows

$$\nabla \cdot \mathbf{E}(\mathbf{r},t) = \rho(\mathbf{r},t)/\epsilon_0 . \quad (62)$$

After properly fixing the electric field $\mathbf{E}(\mathbf{r},t)$ through boundary conditions, the numerical solution of (62) allows one to find $\mathbf{E}(\mathbf{r},t)$ everywhere from the charge density, using Poisson's equation (16), as the field is obtained easily from the scalar potential, as discussed there.

Using (62) in (60) leads to a new, more interesting form

$$\begin{aligned}\frac{\partial \rho(\mathbf{r},t)}{\partial t} + \nabla \cdot J(\mathbf{r},t) &= \\ &= \epsilon_0 \frac{\partial \nabla \cdot \mathbf{E}(\mathbf{r},t)}{\partial t} + \nabla \cdot J(\mathbf{r},t) \quad (63) \\ &= \nabla \cdot \left(\epsilon_0 \frac{\partial \mathbf{E}(\mathbf{r},t)}{\partial t} + J(\mathbf{r},t)\right) = 0 .\end{aligned}$$

The term on the left-hand side of (63) is just the quantum version of the total (particle plus displacement) current $J_T(\mathbf{r},t)$, which is defined through the last term as

$$J_T(\mathbf{r},t) = \epsilon_0 \frac{\partial \mathbf{E}(\mathbf{r},t)}{\partial t} + J(\mathbf{r},t) . \quad (64)$$

If the electromagnetic field is assumed to be "external" and known *a priori*, the numerical solution of (57) is sufficient to compute the total current in (67) through (58) and (61). However, assuming *a priori* knowledge of the electromagnetic fields is not always a good approximation because of the interplay between charges and fields as manifested in (62). A complicated self-consistent solution of the Schrödinger equation (57) along with Maxwell's equations is required. On one side, (57) provides the quantum sources $\rho(\mathbf{r},t)$ and $J(\mathbf{r},t)$, once the vector $\mathbf{A}(\mathbf{r},t)$ and scalar $\varphi(\mathbf{r},t)$ are known. And, on the other side, Maxwell's equations provide the electromagnetic potentials $\mathbf{A}(\mathbf{r},t)$ and $\varphi(\mathbf{r},t)$, once the quantum sources $\rho(\mathbf{r},t)$ and $J(\mathbf{r},t)$ are known. This, of course, means iterating from one equation to the other until self-consistency is achieved.

In some scenarios in the modeling of nano-electronic devices [23], the role of the vector potential $\mathbf{A}(\mathbf{r},t)$ is neglected and the self-consistent loop involves only (57) and (62) (or Poisson's equation). An additional understanding of the physical meaning of the displacement current in (64) can be obtained by recognizing that the terms $\rho(\mathbf{r},t)$ and $J(\mathbf{r},t)$ in (58) and (61), as well as the electric and magnetic fields, refer to results obtained for an ensemble of many repeated experiments.

Now let us return to a description of an individual quantum experiment that can be obtained through the quantum (Bohm) trajectories [22,23,36]. In Bohmian mechanics, the charge density in (58) can be written as an ensemble over many different trajectories

$$\rho(\mathbf{r},t) = e|\Psi(\mathbf{r},t)|^2$$
$$= \frac{1}{M}\sum_{j=1}^{M} e\,\delta(\mathbf{r}-\mathbf{r}^j(t)), \quad (65)$$

where each trajectory $\mathbf{r}^j(t)$ is linked to a particular experiment with $j = 1,2,..,M \to \infty$. Similarly, the ensemble current density in (61) can be rewritten as a sum over different experiments

$$\mathbf{J}(\mathbf{r},t) = \frac{1}{M}\sum_{j=1}^{M} e\,\mathbf{v}(t)\delta(\mathbf{r}-\mathbf{r}^j(t)), \quad (66)$$

where

$$\mathbf{v}^j(t) = \frac{d\mathbf{r}^j(t)}{dt} = \frac{\mathbf{J}(\mathbf{r}^j(t),t)}{|\Psi(\mathbf{r}^j(t),t)|^2} \quad (67)$$

is the Bohmian velocity of the $j$-th trajectory [22]. Notice that all trajectories are guided by the same wave function $\Psi(\mathbf{r},t)$ solution of (57), but different trajectories $\mathbf{r}^j(t) = \mathbf{r}^j(0) + \int_0^t \mathbf{v}^j(t)\,dt$ are generated due to different initial positions $\mathbf{r}^j(0)$. At this point, it is interesting to rewrite the Gauss' law in (62) using (65) as

$$\epsilon_0 \nabla \cdot \mathbf{E}(\mathbf{r},t) = \rho(\mathbf{r},t)$$
$$= \frac{1}{M}\sum_{j=1}^{M}\rho^j(\mathbf{r},t)$$
$$= \frac{1}{M}\sum_{j=1}^{M} e\,\delta(\mathbf{r}-\mathbf{r}^j(t)) \quad (68)$$
$$= \epsilon_0 \nabla \cdot \left(\frac{1}{M}\sum_{j=1}^{M}\mathbf{E}^j(\mathbf{r},t)\right)$$
$$= \frac{1}{M}\sum_{j=1}^{M}\epsilon_0 \nabla \cdot \mathbf{E}^j(\mathbf{r},t) \quad,$$

where the individual $\mathbf{E}^j(\mathbf{r},t)$ are defined as the electric field of an individual experiment satisfying its own individual Gauss' law

$$\varepsilon_0 \nabla \cdot \mathbf{E}^j(\mathbf{r},t) = e\delta(\mathbf{r}-\mathbf{r}^j(t)). \quad (69)$$

Here, an individual particle motion on the RHS defines the *local* charge density of this individual experiment. It becomes now evident from (68) and (69) that the electric field in (62) is a summation of an ensemble of experiments.

It is straightforward to check that a single trajectory $\mathbf{r}^j(t)$, whose charge density is given by the RHS of (69), and whose current is given by the expression within the summation on the RHS of (66), satisfies the continuity equation, as

$$\frac{\partial \rho^j(\mathbf{r},t)}{\partial t} + \nabla \cdot \mathbf{J}^j(\mathbf{r},t) =$$
$$= e\nabla\delta(\mathbf{r}-\mathbf{r}^j(t))\left(-\frac{d\mathbf{r}^j(t)}{dt}\right) + \quad (70)$$
$$+ e\mathbf{v}^j(t)\nabla\delta(\mathbf{r}-\mathbf{r}^j(t)) = 0.$$

This last equation is just a statement that if the electron appears/disappears in one small region of space, a current must appear at the borders of such small region [36]. Then, by using (13) in the left-hand side of (70), one obtains

$$\epsilon_0 \frac{\partial \nabla \cdot \mathbf{E}^j(\mathbf{r},t)}{\partial t} + \nabla \cdot \mathbf{J}^j(\mathbf{r},t) =.$$
$$= \nabla \cdot \left(\epsilon_0 \frac{\partial \mathbf{E}^j(\mathbf{r},t)}{\partial t} + \mathbf{J}^j(\mathbf{r},t)\right) = 0. \quad (71)$$

The terms inside the large parentheses on in the last line of this equation provides a single-particle/experiment form of the total current, including displacement current. This form is free from any divergence terms. This gives the total current due to a single particle/experiment as

$$\mathbf{J}_T^j(\mathbf{r},t) = \epsilon_0 \frac{\partial \mathbf{E}^j(\mathbf{r},t)}{\partial t} + \mathbf{J}^j(\mathbf{r},t), \quad (72)$$

which is the single-particle/experiment form of (64). The divergence-free total current for a single experiment seen in (72) directly implies that the total current computed from the average of an ensemble of experiments seen in (64) is also divergence-free, without the need for the additional discussion at the beginning of this section.

*4.2 Quantum Matter and Quantum Light*

Up until this point, the electromagnetic field has been treated as a classical object. Now, both matter and light will be considered to be quantum entities. The procedure to quantize the electromagnetic field follows the procedure used in quantizing the matter from (55) to (56) [23,44,46]. The first step is introducing the energy of the electromagnetic field $H_R$ into the Hamiltonian (55) as

$$H_{e,p} = \frac{(p-eA)^2}{2m} + e\varphi + H_R , \quad (73)$$

where $H_R$ represents the energy density of the electromagnetic fields

$$H_R = \frac{\epsilon_0}{2} \int d^3r \, (E_\perp \cdot E_\perp + c^2 B \cdot B). \quad (74)$$

The electromagnetic energy depends only on the transverse component $E_\perp$ because the energy assigned to the longitudinal filed $E_\parallel$ is given (in the Coulomb gauge) by the scalar potential $\varphi$, ($E = E_\perp + E_\parallel$) [44,46]. The second step in quantizing light is using the appropriate canonical variables in the descriptions of the electromagnetic fields. Books dealing with quantum optics provide the necessary mathematical tools to find such canonical variables of the electromagnetic fields [44,46]. As an example, we consider a simple scenario of one electron interacting with a single mode electromagnetic field with a unique frequency $\omega$ [23,47,48]. Then, the classical Hamiltonian (73) can be written as

$$\begin{aligned} H_{e,p,I} &= H_e + H_p + H_I \\ &= \frac{1}{2m} p^2 + V(x,t) + \\ &\quad + \frac{\omega}{2}(q^2 + s^2) - \frac{\alpha}{\sqrt{\hbar}} xq \, . \end{aligned} \quad (75)$$

where the first term on the third line corresponds to the energy of the electromagnetic field in (73) written in terms of the canonical variables $q$ and $s$ [47]. The Hamiltonian in (75) is based on assuming that the wavelength of the electromagnetic field is much larger than the typical spatial region where the electrons move. In other words, the spatial dependence of the vector potential is ignored [46]. The parameter $\alpha$ controls the strength of the light-matter interaction [47]. For the case $\alpha = 0$, the Hamilton equations of motion for the description of the light, when applied to the Hamiltonian $H_{e,p,I}$ (75), are

$$\begin{aligned} \frac{dq}{dt} &= \frac{\partial H_{e,p,I}}{\partial s} = \omega s \\ \frac{ds}{dt} &= -\frac{\partial H_{e,p,I}}{\partial q} = -\omega q \end{aligned} . \quad (76)$$

Combining these two equations, one can find the equation of motion for $q$ as

$$\frac{d^2 q}{dt^2} = -\omega^2 q. \quad (77)$$

Clearly the solution of (77) gives a sinusoidal signal, $q(t) \propto \sin(\omega t)$, which represents the expected time-dependence of the free single-mode electromagnetic field.

The third step to reach quantum light is the canonical quantization of the canonical variables following the prescription just above (56). In terms of the variables used above for light, this leads to the substitutions

$$\begin{aligned} q &\to \sqrt{\hbar} q \\ s &\to -i\sqrt{\hbar} \frac{\partial}{\partial q} \end{aligned} , \quad (78)$$

so that $[q_i, s_j] = i\hbar \delta_{ij}$. Finally, introducing these substitutions into the classical Hamiltonian (75), the new full quantum Hamiltonian is [23,47]

$$\begin{aligned} H_{e,p,I} &= H_e + H_p + H_I \\ &= \frac{\hbar^2}{2m_0} \frac{\partial^2}{\partial x^2} + V(x,t) \\ &\quad -\frac{\hbar\omega}{2} \frac{\partial^2}{\partial q^2} + \frac{\hbar\omega}{2} q^2 - \alpha x q \end{aligned} . \quad (79)$$

The quantum state of light and matter is now described by the time-dependent wave function $\Psi(x, q, t)$ that contains information of the electron through the variable $x$ and information of the light through $q$. Such a wave function is a solution of the following Schrödinger equation

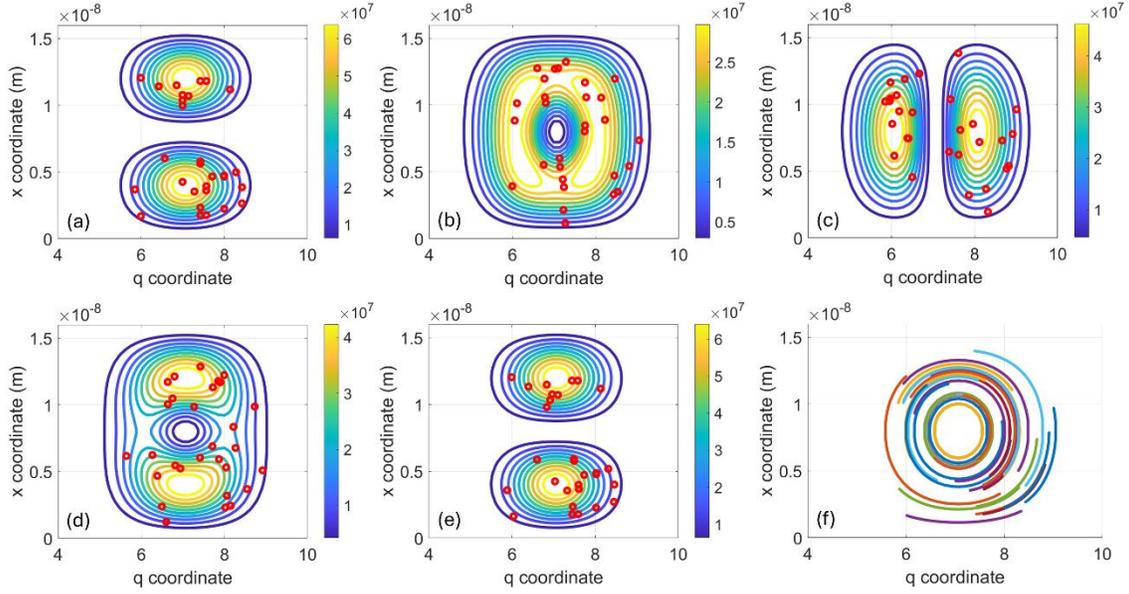

**Figure 7** Probability density of the wavefunction $\Psi(x,q,t)$ in the $x-q$ configuration space at $t$=0 fs (a), $t$=20 fs (b), $t$=40 fs (c), $t$=60 fs (d) and $t$=80 fs (e). Red circles indicate the positions of $M=30$ Bohmian trajectories $x^j(t)$ and $q^j(t)$ selected with random initial positions according to $|\Psi(x,q,0)|^2$. In (f), the continuous path of these trajectories is plotted, showing their (Rabi) oscillations.

$$i\hbar \frac{\partial \Psi(x,q,t)}{\partial t} = \left(-\frac{\hbar^2}{2m_0}\frac{\partial^2}{\partial x^2} + V(x,t) - \frac{\hbar\omega}{2}\frac{\partial^2}{\partial q^2} + \frac{\hbar\omega}{2}q^2 - \alpha x\, q\right)\Psi(x,q,t) \quad .(80)$$

From a mathematical point of view, (80) is just a 2D Schrödinger equation whose numerical solution can be worked out in a straight forward manner.

The behavior of $\Psi(x,q,t)$ can be easily anticipated by using the product wave function $\psi_{e,i}(x)\psi_{p,j}(q)$, where $\psi_{e,i}(x)$ is defined as the basis function of the electron inside an infinite well described by the Hamiltonian

$$H_e = -\frac{\hbar^2}{2m_0}\frac{\partial^2}{\partial x^2} + V(x,t) \quad (81)$$

in (80) and $\psi_{p,j}(q)$ is the basis function of the light described by the quantum harmonic oscillator of the Hamiltonian

$$H_p - \frac{\hbar\omega}{2}\frac{\partial^2}{\partial q^2} + \frac{\hbar\omega}{2}q^2 \quad (82)$$

in (80). In particular, $\psi_{e,0}(x)$ and $\psi_{e,1}(x)$ are the two first electron eigen-states with eigen-energies $E_{e,0}$ and $E_{e,1}$, and $\psi_{p,0}(q)$ and $\psi_{p,1}(q)$ are the two eigen-states of the light with eigen-energies $E_{p,0} = \hbar\omega/2$ and $E_{p,1} = 3\hbar\omega/2$. The energies of both the electrons and the light are quantized. In particular, when the light is described by $E_{p,1} = 3\hbar\omega/2$, we say that the light has one photon. When the light has the minimum energy value $E_{p,0} = \hbar\omega/2$, we say that the light has no photons (vacuum state [44][46]).

Now, the above description may seem to be somewhat simple, but it corresponds to a standard model in both quantum optics [49] and quantum information [50]. This model is the Jaynes-Cummings model [51]. The model considers the coupling of a two (energy) level atom, which is our "matter" and a mode in an optical cavity which is described as a harmonic oscillator, with quantized energy levels, which is our "light." A weak interaction between these two systems, the light-matter

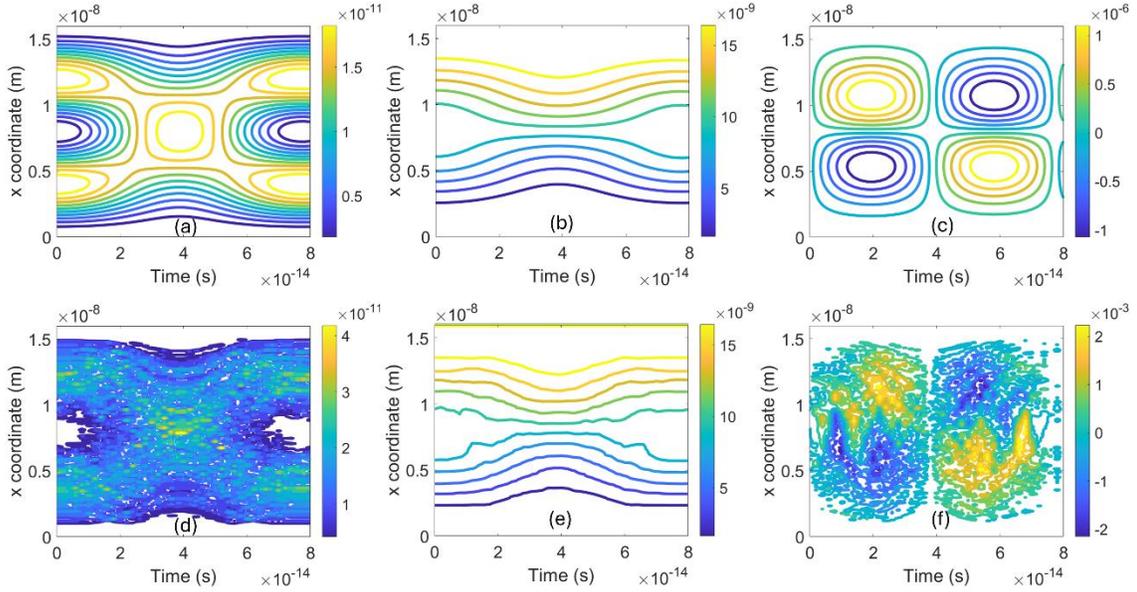

**Figure 8** (a) Ensemble "Charge" density obtained from $\rho(x,t) = \int |\Psi(x,q,t)|^2 dq$. in Fig. 1. (b) Ensemble electric field from Gauss's law applied to $\rho(x,t)$ in (a). (c) Ensemble displacement current as the time-derivative of the electric field in (b). (d) Ensemble "Charge" density obtained a sum of the single-experiment charge densities $\rho^j(x,t) = e\delta(x - x_j(t))$ from different J=1,2,…,1000 Bohmian trajectories (some of them depicted in Fig. 1). (e) Ensemble electric field computed as a sum of the single-experiment electric fields solution of the Gauss law $\varepsilon\, \partial E^j(x,t)/\partial x = \rho^j(x,t)$. (f) Ensemble displacement current computed as a sum of individual displacement currents computed from the time derivative of $E^j(x,t)$.

interaction, causes the energy to move back and forth between the atom and the cavity. This is just like two coupled pendula, where the oscillation itself oscillates between the two pendula. This motion of energy back and forth is often referred to as a Rabi oscillation, although this term is usually assigned to the amplitude oscillation of the two-level atom [49].

This behavior is illustrated in Fig. 7. In the contour plots in panel (a) of Fig. 7, the initial state given by $\Psi(x,q,0) \approx \psi_{e,1}(x)\psi_{p,0}(q)$ with total energy $E_{e,1} + E_{p,0}$ is plotted. Notice that the probability of this state has two maxima in the $x$ direction (corresponding to $\psi_{e,1}(x)$) and one maximum in the $q$ direction (corresponding to $\psi_{p,0}(q)$). Under the appropriate resonant conditions for light and matter given by $E_{e,1} - E_{e,0} \approx E_{p,1} - E_{p,0}$, another hybrid state arises that has the same energy as the initial one. In panel (c) of Fig. 7, this new state is $\psi_{e,0}(x)\psi_{p,1}(q)$ which has energy $E_{e,1} + E_{p,0} \approx E_{e,0} + E_{p,1}$. The probability distribution of this new state now has only one maximum in the $x$ direction (corresponding to $\psi_{e,0}(x)$) and two maximums in the $q$ direction (corresponding to $\psi_{p,1}(q)$). Thus, the wave function $\Psi(x,q,t)$ solution of (73) oscillates between $\psi_{e,1}(x)\psi_{p,0}(q)$ (panel (a) and (e) of Fig. 7) and $\psi_{e,0}(x)\psi_{p,1}(q)$ (panel (c) in Fig. 7) at the Rabi frequency [47].

Again, in any discussion of displacement current, it becomes relevant to separate the properties of an ensemble of experiments from the properties of an individual experiment. We turn now to the former case.

### 4.3 Displacement Current for an Ensemble of Experiments

The electron charge density can be obtained from the modulus of the wave function solution of (80) in the form

$$\rho(x,t) = \int_{-\infty}^{\infty} e|\Psi(x,q,t)|^2 dq, \quad (83)$$

which must be interpreted as the probability of finding the electron at position $x$ at time $t$ when the same experiment is repeated many times, for whatever property $q$ of the light. The unitary evolution of $\Psi(x,q,t)$ in (80) ensures that there is a global conservation of probability in the whole system. In fact, a straightforward manipulation [47] of (80) similar to what has been done in (60), shows that the total probability is conserved because it satisfies the following (local) continuity equation

$$\frac{\partial |\Psi|^2}{\partial t} + \frac{\partial J_e}{\partial x} + \frac{\partial J_p}{\partial q} = 0, \quad (84)$$

where $J_e(x,q,t)$ is the particle current of the electron and $J_p(x,q,t)$ is the "current" of the amplitude $q(t)$ of the electromagnetic mode. The continuity equation in the configuration space in (84) can be transformed into a continuity equation in the physical space by disregarding the details of the light as done in (83) by integrating the equation

$$\int_{-\infty}^{\infty} \frac{\partial |\Psi|^2}{\partial t} dq + \int_{-\infty}^{\infty} \frac{\partial J_e}{\partial x} dq$$
$$+ \int_{-\infty}^{\infty} \frac{\partial J_p}{\partial q} dq = \frac{\partial \rho(x,t)}{\partial t} + \frac{\partial J(x,t)}{\partial x} \quad (85)$$
$$= 0,$$

where the last term on the LHS vanishes since $J_e(x,\pm\infty) = 0$ and the electron current density is defined as

$$J(x,t) = \int_{-\infty}^{\infty} J_e(x,q,t) dq. \quad (86)$$

The charge density in (83) and current density in (86) satisfy the continuity equation defined in the right-hand side of (85). Thus, all the ingredients to reach a divergenceless total (particle and displacement) current are obtained. The discussions that follow (62) to (64) in section 4.1 define the quantum version of the total current and this can be identically reproduced here.

As an example, the charge density, electric field, and displacement current are plotted in panels (a)-(c) of Fig. 8, respectively. They are computed from the same wave function $\Psi(x,q,t)$ whose time evolution is shown in Fig. 7.

Since the light description is included inside the Hamiltonian (73), there is no need for a self-consistent solution of the Schrödinger equation (82) and Maxwell equations, as had happened in the classical treatment of the light in Sec. 4.1. As seen in (80), the electromagnetic fields are already defined from the Hamilton equations applied to the Hamiltonian $H_{e,p,I}$ in (75).

### 4.4 Displacement Current for a Single Experiment

The description of quantum phenomena in terms of Bohmian trajectories that has been developed above now allows us to describe an individual light-matter experiment [23,47]. One can rewrite (83) here as

$$\rho(x,q,t) = e|\Psi(x,q,t)|^2$$
$$= \frac{1}{M}\sum_{j=1}^{M} e\,\delta(x - x^j(t)) \quad (87)$$
$$\times \delta(q - q^j(t)).$$

Here, the $x^j(t)$ is the quantum (Bohmian) trajectory of an electron guided by the wave function solution of (82) following

$$x^j(t) = x^j(0) + \int_0^t v_e^{\,j}(t)\,dt, \quad (88a)$$

with

$$v_e^{\,j}(t) = \frac{dx^j(t)}{dt} = \frac{J_e(x^j(t),q^j(t),t)}{|\Psi(x^j(t),q^j(t),t)|^2} \quad (88b)$$

the Bohmian velocity of the trajectory [23,47]. Identically, $q^j(t)$ is the trajectory of the parameter $q$ that defines the amplitude of the electromagnetic field as a function of time

$$q^j(t) = q^j(0) + \int_0^t v_p^{\,j}(t)\,dt \quad (89a)$$

with

$$v_e^j(t) = \frac{dx^j(t)}{dt} = \frac{J_e(x^j(t),q^j(t),t)}{|\Psi(x^j(t),q^j(t),t)|^2} .(89b)$$

the Bohmian velocities determining how fast the amplitude of the electromagnetic field changes with time [23,47].

The red circles in Fig. 7 show the evolution of the Bohmian trajectories for the electron and light $\{x^j(t), q^j(t)\}$ in the same configuration space as the wave function. They perfectly reproduce the evolution of $|\Psi(x,q,t)|^2$ at all times.

Using (87), the charge density (83) can be written here as

$$\rho(x,t) = \int_{-\infty}^{\infty} e|\Psi(x,q,t)|^2 dq = \frac{1}{M}\sum_{j=1}^{M} e\,\delta(x - x^j(t)). \quad (90)$$

The use of (83), and the continuity equation in the right-hand side of (85), allows one to straight-forwardly reproduce the definitions of the electric field and the displacement current of an individual experiment done in (68)-(72) in section 4.1.

The charge density is plotted In Fig. 8(d) using a 1D version of (90) as a sum of $M = 1000$ individual densities $\rho^j(x,t) = e\delta(x - x^j(t))$. This charge density of Fig. 8(d) exactly reproduces the charge density plotted in Fig. 8(a) computed from only the wave function. The dynamics of the electrons essentially involve a transition from being localized at two maxima at the borders of the infinite well (given by the state $\psi_{e,1}(x)$) to being localized around one maximum at the center (given by the state $\psi_{e,0}(x)$), and *vice versa*. Since the charge density changes with time, the electric field will also have a time dependence, as seen in Figs. 8(b) and 8(e). This time-dependent electric field generates the displacement current shown in Fig. 8(c) and 8(f). There is great agreement between the average-over-ensemble of experiments results computed from the wave function (top panels in Fig. 8) and those computed from $M = 1000$ trajectories (bottom panels in Fig. 2). Note that the computations based on trajectories are noisier and therefore on needs more trajectories to achieve distributions as smooth as the ones computed from the wave function.

## 5. Discussion

By now, it should be clear to the reader that the thread running through this work is the importance of displacement current, in that it forces the consideration of time-varying events into electromagnetics. Certainly, nearly everyone (in physics and electrical engineering) is trained in Maxwell's equations and electromagnetic waves, but it is seldom that they realize the role played by displacement current, particularly in lower frequency circuits. Without this time varying term, there would be no wave equations for use in fields ranging from electric power distribution to optical information processing. The time varying term can be important over a range of frequencies, extending from tens of Hz to THz.

There is also an important caveat that comes with the importance of time variation. Phrases, even when almost universally accepted, such as "dielectric constant" are an oxymoron. The dielectric function is never constant except over very narrow frequency ranges--and the circumstances in which it can be considered constant forms a very small set of conditions that are usually very special.

Even in what is known as linear response [52], the dielectric function of a simple material like a semiconductor is a very complicated (and even nonlinear) function of frequency with multiple poles and zeroes [53], and it is further complicated by the formation of excitons (bound electron and hole pairs), the existence of band-gap narrowing, and other dynamic effects. The method of studying this dielectric function is spectroscopy [54]. In composite systems, this becomes much harder to accomplish.

It is clear that a simple system such as the ion channel of Figure 5 is an enormously complicated compound system, with each atom or molecular structure having its own dielectric response. The properties of ion channels change dramatically when as few as a handful of atoms are changed and so dielectric properties must be understood with atomic resolution [55]. Determining the overall dielectric response is extremely difficult and challenges our level of understanding at the fundamental level [13, 56].

In modern semiconductor devices, layers of thin film materials are stacked and adjoined to one another. Even with simple stacking of thin films, such as in growth of superlattices and heterostructures, determination of the dielectric response, even over a limited range of frequency, is difficult [57]. By the time one tries to couple single photons to single quantum dots embedded into photonic bandgap material, the task is almost impossible [58]. Even the optical dielectric response of a single semiconductor (or even metal) layer is governed by the valence (bonding) electron response to the a.c. signals, and this is usually in the ultra-violet spectral region. Determination of the temporal response in this region, and the delay in which the electrons follow the optical signal, has been determined by the use of attosecond laser pulses [59].

Sectio 4 demonstrated that there is no fundamental difference between the classical and quantum descriptions of the displacement current [23,33]. If needed, both can be described in terms of individual trajectories or at the level of an ensemble over many experiments. The only differences between quantum and classical displacement currents are that the dynamics of quantum electrons exhibit exotic behaviors which are not present in classical dynamics, such as tunneling and energy quantization [23,33,60]. Apart from this, everything that we have discussed in previous sections for classical systems applies to quantum ones as well.

The use of individual Bohmian trajectories has another advantage that is typically unnoticed in the literature. As seen in Fig. 7(f), each electron in a single experiment oscillates, but the ensemble results tend to wash out these oscillations because some electrons oscillate from right to left in some experiments, while others from left to right in other experiments. But, in the laboratory, only one experiment at a time is conducted. Thus, in some scenarios, the information from a single experiment can be more relevant than that obtained from an ensemble of experiments. The same happens, for example, when dealing with electronic devices at high frequencies where displacement current is fully relevant [23,60]. The proper behavior of the displacement current in an individual electron device is not guaranteed by the fact that a displacement current computed as an average over an ensemble of identical devices is adequate. In other words, if you put one hand in the fridge and the other hand on the oven, the ensemble temperature of your hands might seem satisfactory, but neither of your hands will work satisfactorily!

There is a beautiful moral that arises from the discussion of this section. It seems that Kirchhoff was motivated by the continuity equation to postulate the origin of the displacement current. Despite the fact that the concept of an electron was not known at that time, the meaning of Kirchhoff's law (4), or the continuity equation (6), indicates that the electrons leaving a volume are equal to the those entering plus the time change of the electron density inside the volume. By Gauss' law, such temporal variation of the charge generates displacement current. Kirchhoff's intuition at the middle of the 19$^{th}$ century is the seminal work for the development of our information society, which actually may be considered to extend from the telegraph in the 1840s to today's internet. After almost two centuries, his intuition is still helpful in computing displacement current, even in modern quantum electron devices working at THz frequencies, allowing the never-ending progress of our information society. Indeed, addition of the displacement current to Ampere's Law, and the continuity of the total current, provide the important elements necessary to design and understand our circuits for today's technology and provide striking insights into many other systems of daunting complexity.

There is an additional lesson extracted from the discussion of the quantum displacement current. A quantum description of a single experiment can be done in terms of Bohmian trajectories. A single trajectory satisfies a continuity equation by construction as seen in (70). If the electron appears/disappears in one small region, a current must appear at one of the borders of such small region. From such a trivial continuity equation for a single trajectory, a divergence-less total current for a single-experiment can be easily deduced using Gauss' law, as seen in (72). Then, an ensemble of trajectories/experiments also imply a divergence-free total current, by construction. At this point one can interpret that (64) is just a consequence of (72), meaning that the fundamental origin of the divergence-less total current is just the fact that matter, even at the quantum level, can be described by quantum trajectories. Of course, one is also allowed to interpret that (64) is more fundamental than (72) and that (72) is just a byproduct of (64) with no deeper ontological implications. In this second option, the fundamental origin of the divergence-free total current is hidden in the fact that the Hamiltonians used to describe nature include a continuity equation inside (as will be pointed out in the Appendix). Thus, displacement current is in fact just a fundamental law of nature.

**Appendix. The Continuity Equation and Gauge Invariance**

Throughout the paper, it has been emphasized at several points just how the displacement current emerges naturally from a continuity equation. In this appendix, we discuss how the continuity equation can be understood as a mandatory needed in order to construct a gauge theory for (classical or quantum) electromagnetics. Combining both results, we conclude that the displacement current is a fundamental ingredient of any physical theory.

In the paper, the interaction of light and matter is described through the Hamiltonian $H$ defined in (55). Here, it is more convenient to discuss the interaction of "light" and "matter" through the Lagrangian $L$ defined as

$$\begin{aligned} L &= \boldsymbol{v} \cdot \boldsymbol{p} - H. \\ &= \boldsymbol{v} \cdot \boldsymbol{p} - \frac{(\boldsymbol{p}-e\boldsymbol{A})^2}{2m} - e\varphi \\ &= m\boldsymbol{v}^2 + e\boldsymbol{v} \cdot \boldsymbol{A}. \quad (A1) \\ &\quad -\frac{1}{2}m\boldsymbol{v}^2 - e\varphi \\ &= \frac{1}{2}m\boldsymbol{v}^2 + L_I \end{aligned}$$

where we have used $m\boldsymbol{v} = \boldsymbol{p} - e\boldsymbol{A}$, which gives the definition of the velocity. Thus, in this framework, the interaction of "light" and "matter" is given by the interacting Lagrangian $L_I = e\boldsymbol{v} \cdot \boldsymbol{A} - e\varphi$. In fact, the term $e\boldsymbol{v}$ can be understood as the conduction current $\boldsymbol{J}(\boldsymbol{r},t) = e\boldsymbol{v}\delta(\boldsymbol{r} - \boldsymbol{r}_a(t))$ evaluated at the position $\boldsymbol{r}_a(t)$ of the particle, while $e$ is the charge density $\rho(\boldsymbol{r},t) = e\delta(\boldsymbol{r} - \boldsymbol{r}_a(t))$ at the position of the particle. Thus, the coupling between matter and light is defined by the Lagrangian $L_I$ written as

$$L_I = e\boldsymbol{v} \cdot \boldsymbol{A} - e\varphi.$$
$$= \int_{-\infty}^{\infty} d\boldsymbol{r}^3 [\rho(\boldsymbol{r},t)\varphi(\boldsymbol{r},t) \\ -\boldsymbol{J}(\boldsymbol{r},t) \cdot \boldsymbol{A}(\boldsymbol{r},t)] \quad . \quad (A2)$$

The development leading to (A.2) has been done for a simple single-particle system, but the final expression in this equation is completely general either for classical or quantum systems.

Hamilton's principle (also known as the stationary-action principle or the principle of least action) says that the physical trajectories $\boldsymbol{r}(t)$ of the system from the initial time $t_0$ to the final time $t_f$ are the ones that are stationary points of the system's action functional constructed from the Lagrangian as

$$S(\boldsymbol{r}(t),t) = \int_{t_0}^{t_f} dt \, L(\boldsymbol{r}(t),t) . \quad (A3)$$

The Euler-Lagrange equations are the equations of motion of the system, written in terms of the Lagrangian, that minimize the action. Such equations of motions must be gauge invariant. Thus, when putting (A.2) into (A.3), one obtains

$$S_I = \int_{t_0}^{t_f} dt \, L_I \\ = \int_{t_0}^{t_f} dt \int_{-\infty}^{\infty} d\boldsymbol{r}^3 [\rho(\boldsymbol{r},t)\varphi(\boldsymbol{r},t) \\ -\boldsymbol{J}(\boldsymbol{r},t) \cdot \boldsymbol{A}(\boldsymbol{r},t)] \quad (A4)$$

that has to lead to gauge invariant Euler-Lagrange equations. As mentioned in equation (21) of the paper, the electromagnetic potentials $\varphi(\boldsymbol{r},t)$ and $\boldsymbol{A}(\boldsymbol{r},t)$ inside (A.4) are gauge dependent, but the Euler-Lagrange equation generated from (A.4) must be gauge invariant. In order to see how this happens, one defines $S_I^\Lambda$ as the expression of the interacting Langrangian (A.4) in another gauge $\Lambda$ as

$$S_I^\Lambda = -\int_{t_0}^{t_f} dt \int_{-\infty}^{\infty} d\boldsymbol{r}^3 [\rho(\boldsymbol{r},t) \\ \times \left(\varphi(\boldsymbol{r},t) - \frac{\partial \Lambda(\boldsymbol{r},t)}{\partial t}\right) \\ -\boldsymbol{J}(\boldsymbol{r},t) \cdot \left(\boldsymbol{A}(\boldsymbol{r},t) + \boldsymbol{\nabla}\Lambda(\boldsymbol{r},t)\right)] \quad (A5)$$

which leads to

$$S_I^\Lambda = S_I \\ + \int_{t_0}^{t_f} dt \int_{-\infty}^{\infty} d\boldsymbol{r}^3 \left[\rho(\boldsymbol{r},t)\frac{\partial \Lambda(\boldsymbol{r},t)}{\partial t} \right. \\ \left. +\boldsymbol{J}(\boldsymbol{r},t) \cdot \boldsymbol{\nabla}\Lambda(\boldsymbol{r},t)\right] . \quad (A6)$$

The integration by parts of the last terms in this equation lead to

$$S_I^\Lambda = S_I + \int_{t_0}^{t_f} dt \int_{-\infty}^{\infty} d\boldsymbol{r}^3 \Lambda(\boldsymbol{r},t) \\ \times \left(\frac{\partial \rho(\boldsymbol{r},t)}{\partial t} + \boldsymbol{\nabla} \cdot \boldsymbol{J}(\boldsymbol{r},t)\right) . \quad (A7)$$

where it has been used that

$$\int_{-\infty}^{\infty} d\boldsymbol{r}^3 \boldsymbol{\nabla}\bigl(\boldsymbol{J}(\boldsymbol{r},t)\Lambda(\boldsymbol{r},t)\bigr) = 0 \quad (A8)$$

because $\boldsymbol{J}(\boldsymbol{r},t) = 0$ at $|\boldsymbol{r}| \to \infty$, and it can be shown that

$$\int_{t_0}^{t_f} dt \frac{\partial}{\partial t}\bigl(\rho(\boldsymbol{r},t)\Lambda(\boldsymbol{r},t)\bigr) \quad (A9)$$

does not alter the Euler-Lagrange equation because this term only depends on the initial and final times (not on the path).

Throughout the paper, the natural connection between the continuity equation and the displacement current has been discussed (see (6) or (60)). Here, we have shown that a general physical theory that is gauge invariant must satisfy the continuity equation (A.7). The overall conclusion is that the displacement current may be thought of as being at the origin of gauge theories.

**References**

[1] Maxwell J C 1855 *Trans. Cambridge Phil. Soc.* **10** 155
[2] Shockley W 1956 *Electrons and holes in semiconductors, with applications to transistor electronics* (New York: Krieger)
[3] Maxwell J C 1861 *Phil. Mag., Ser. 4* **32** 11
[4] Kirchhoff G 1857 *Ann. Phys. Pogg*. **102** 193; Tr. Kirchhoff G 1857 *Phil. Mag.*, Ser. 4 **13** 393
[5] Graneau P and Assis A K T 1994 *Apeiron* **19** 19
[6] Faraday M 1852 *Phil. Mag., Ser. 4* **3** 401
[7] Landau L 1930 *Z. Phys*. **64** 629
[8] Rabi I I 1928 *Z. Phys*. **49** 507
[9] Bertoncini R and Jauho A-P 1992 *Semicond. Sci. Technol.* **7** 8
[10] Bertoncini R and Jauho A-P 1992 *Phys. Rev. Lett*. **68** 2826
[11] Lynch W T 1972 *J. Appl. Phys*. **53** 3274
[12] Barsukov E and MacDonald J R 2018 *Impedance Spectroscopy: Theory, experiments, and applications* (New York, John Wiley)
[13] Leuthold J, Koos C and Freude W 2010 *Nature Photon*. **4** 535
[14] Ferry D K and Fannin R D 1971 *Physical Electronics* (Reading, MA: Addison-Wesley) pp. 143-145
[15] Child C D 1911 *Phys. Rev. (Ser. 1)* **32** 492
[16] Langmuir I 1913 *Phys. Rev*. **2** 450
[17] Ferry D K 2021 *Quantum Mechanics: An Introduction for Device Physicists and Electrical Engineers* (Boca Raton: CRC Press)
[18] Cooper F, Khareb A and Sukhatme U 1995 *Phys. Repts*. **251** 267
[19] Goldstein H 1950 *Classical Mechanics* (Reading, MA: Addison-Wesley)
[20] Feynman R P 1948 *Rev. Mod. Phys*. **20** 367
[21] Kennard E H 1928 *Phys. Rev*. **31** 876
[22] Bohm D 1952 *Phys. Rev*. **85** 166
[23] Ferry D K, Oriols X and Weinbub J 2023 *Quantum Transport in Semiconductor Devices: Simulation Using Particles* (Bristol, UK: IOP Publishing)
[24] de Broglie L 1923 *Nature* **112** 540
[25] M. Kline and I. W. Kay 1965 *Electromagnetic Theory and Geometrical Optics* (New York: John Wiley)
[26] V. Radisic *et al*. 2012 *IEEE Transactions on Microwave Theory and Techniques* **60** 724
[27] D. Guerra *et al*. 2012 *2012 IEEE/MTT-S International Microwave Symposium Digest* (New York: IEEE Press) 1-3
[28] Hodgkin A L and Keynes R D 1955 *J. Physiol.* **128** 61
[29] Hille B 2001 *Ion Channels of Excitable Membranes* (Oxford: Oxford University Press)
[30] Hodgkin A L and Huxley A F 1952 *J. Physiol.* **117** 500
[31] Mirenenko A *et al.* 2021 *J. Mol. Bio.* **433** 167002.
[32] Landauer R 1992 *Phys. Scripta* **T42** 110
[33] Oriols X and Ferry D K 2021 *Proc. IEEE* **109** 955.
[34] Oriols X and Ferry D K *2013 J. Comp.Electron*. **12** 317
[35] Bohm D 1952 *Phys. Rev*. **85** 166
[36] Oriols X and Mompart J 2019 *Applied Bohmian Mechanics: from naoscale systems to cosmology,* 2nd edition, (Singapore: Jenny Stanford Publishing)
[37] Oriols X 2007 *Phys. Rev. Lett*. **98** 066803
[38] Marian D, Zanghi N and Oriols X 2016 *Phys. Rev. Lett*. **116** 110404
[39] Villani M.*et al*. 2021 *IEEE Electron Dev. Lett*. **42** 224
[40] Schwinger J 1951 *Proc. Nat. Acad. Sci*. **37** 452
[41] Martinez A *et al*. 2016 *J. Comp. Electron*. **15** 1130)
[42] Barker J R 2010 *J. Comp. Electron*. **9** 243
[43] Ferry D K 2023 *Semicond. Sci. Technol*. **38** 055005
[44] Cohen-Tannoudji C *et al.* 2004 *Photons and Atoms: Introduction to Quantum Electrodynamics* (New York: Wiley).
[45] Eisenberg B, Oriols X and Ferry D 2017 *Molecular Based Mathematical Biology* **5** 78
[46] Grynberg G *et al.* 2011 *Introduction to Quantum Optics: From the Semi-Classical Approach to Quantized Light* (Cambridge: Cambridge University Press)
[47] Destefani C *et al.* 2022 *Phys. Rev B* **106** 205306
[48] Villani M *et al*. 2021 *J. Comp. Elect*. **20** 2232
[49] Birrittella R, Chang K and Gerry C C 2015 *Opt. Commun*. **354** 286
[50] Ferry D K 2024 *Quantum Information in the Nanoelectronic World* (New York: Springer)
[51] Jaynes E T and Cummings F W 1963 *Proc. IEEE* **51** 89
[52] Lindhard J 1954 *Danske Matematisk-Fysiske Meddelelser* **28** 1)
[53] Ferry D K 1991 *Semiconductors* (New York: Macmillan) Ch. 12
[54] Barsoukov E and Ross Macdonald J 2018 *Impedance Spectroscopy: Theory, Experiment, and Applications* (New York: Wiley)

[55] Catacuzzeno L et al. 2021 *J. Gen. Physiol.* **153** e202012706
[56] Demchenko A P 2013 *Ultraviolet Spectroscopy of Proteins* (Berlin: Springer)
[57] Sohie G R L and Maracas G N 1994 *Proc. IEEE Conf. Control Appl.* **1** 539
[58] Lodahl P, Mamoodian S and Stobbe S 2015 *Rev. Mod. Phys.* **87** 347
[59] Hassan M Th et al. 2016 *Nature* **530** 66
[60] Villani M et al. 2021 IEEE Electron Dev. Lett. **42** 224